\documentclass{article}
\usepackage[utf8]{inputenc}
\usepackage{indentfirst} 

\title{
  Botnets Drilling Away Privacy Infrastructure\\
}

\author{
  Yang Yang\\
  \texttt{kyang@ccs.neu.edu}
  \and
  Christophe Leung\\
  \texttt{tophe@ccs.neu.edu}
}
\date{December 2015}

\begin{document}
\maketitle

\section{Abstract}
In this paper, we explore various technologies and their roles in subverting the privacy infrastructure of the Internet. We also provide mitigation techniques on the attack vectors the technologies provide, and assess the overall severity of these threats.

\section{Introduction}
Over the last two decades, advances in privacy-enhancing technologies, including cryptographic mechanisms, standardized security protocols, and infrastructure, significantly improved the privacy of users. Cryptographic primitives are now commonly used in the development of applications, where protocols such as TLS/SSL are widely used to secure web access, VPN tunnels, and wireless networks (e.g., WPA-Enterprise). In the last decade, Tor, a byproduct of those primitives, emerged as a practical solution to protecting the privacy of citizens against censorship and tracking. At the same time, Tor's success encouraged illegal activities, including sophisticated botnets, ransomware, and a marketplace for drugs and contraband. These botnets have proved increasingly adaptable and resistant to various countermeasures that researchers have constructed. One kind of botnet, called OnionBots, leverages the Tor network directly to infect hosts. In Section 3, we will go into more detail about a botnet. Section 4 will delve deeper into the specifics of Tor. With this information, in Section 5, we will explore OnionBots and how they work. We also discuss the threat severity and mitigation tactics for each of these sections. Finally, we touch upon

\section{Botnets Strengthening Capabilities}
A botnet is a plurality of Internet computers that have been set up to forward transmissions to other computers on the Internet. A computer part of a botnet may or may not be aware of the fact that it is obeying the commands given by a central computer, called the master. This master, who issues commands called Command and Control messages, rallies all the computers to accomplish a task, most often for a destructive purpose. This can include sending spam, instigating click-fraud, launching distributed denial-of-service attacks, and spreading malicious software such as adware, spyware, worms, and viruses.~\cite{gomez1993taxonomyBotnet} Botnets can have different variants - centralized, peer-to-peer, and hybrid. Having a centralized server initiating all the attacks results in faster convergence and propagation and is easier to maintain and monitor. However, it also has a single point of failure - If the command server is compromised, the entire botnet is incapacitated.
Botnets operate most efficiently when undiscovered, so attackers organizing botnets are constantly searching for new ways to evade detection. This is why Tor is so attractive for botnets, which we will discuss later on. Some ways to avoid detection without using Tor include using Fast flux and Domain Name Generation. In Fast flux, a botnet maps many IP addresses to a single domain name. It then rapidly registers and deregisters these domain names with the various IPs through changing the DNS A Record, making it tough to detect the actual source of the attack. In Domain Name Generation, new domain names are created for the botnet nodes to use so that they can contact their master.

To mitigate botnets, we have to detect the nodes themselves.~\cite{gu2008botminer}~\cite{moshchuk2009spammingBotnets} Once found, we can hijack and shut down their command and control servers~\cite{perdisci2015malwareBehavior} through a number of different methods. Some include reverse engineering the algorithm used for the Domain Name Generation stated above, and block access to the domains before they can do heavy damage. We can also inspect the network traffic and identify the key features of Command and Control communication. This can be done through machine learning to detect flow sizes and client access patterns~\cite{yadav2010algorithmicallyGeneratedDN}.

On the other hand, the increasing use of Tor by botnets can effectively hide the botnet nodes, making it extremely hard to detect, hijack and shut down. The severity of botnets as a threat to users has remained consistent throughout its inception. Indeed, there are many companies whose entire business model relies on helping companies protect themselves from botnet attacks. Research in this area has also garnered a significant amount of attention. However, attackers are constantly evolving as well, coming up with new ways to circumvent server patches and various other techniques for combating botnets. As we'll see with Tor and OnionBots, the anonymous nature of the nodes creates a whole new playing field for botnets in network security.

\section{Tor, An Anonymity Network}
Tor is a distributed, low-latency anonymity network where clients establish anonymous communication by relaying traffic through intermediary nodes. These intermediary nodes, called onion routers, are Tor relays that pass on the encrypted information from source to destination. The client will package its data and send it to these relays together with symmetric key negotiation to form a circuit. This relay is called the introduction point, which is randomly chosen. It encrypts its message in layers so that each relay can unwrap the top layer with its private key, and pass on the information to the next layer. This repeats until we reach the rendezvous point, which then forwards the data to the intended server.

Tor is excellent at maintaining the anonymous identity of the sender. Each relay only knows where it should send the data next, so it's impossible to backtrack its chain of relays to the original sender. As a result, this identity protection creates a very effective medium for illegal activity. Tor allows users to host Internet servers without revealing their location, which has resulted in technologies such as Silk Road~\cite{christin2012silkroad}, Zeus botnet~\cite{tarakanov2013zeus}, and the hosting of the Cryptolocker Ransomware's C\&C server. ~\cite{keith2013cryptolocker}

Tor itself does not pose a threat to security protocols, as it is just a medium to distributing anonymous information. That being said, the implications of anonymity-focused communication makes it a very potent platform for antagonistic software to be built upon. Botnets can now communicate via Tor and make it extremely difficult to be discovered and torn down, as we will see in the next section. With its increasing popularity, Tor will cause internet security headaches for years to come.

\section{OnionBot, A Botnet Utilizing Tor}
OnionBot is a peer-to-peer botnet that relies on Tor for its communication with each other~\cite{2012databreach}. Because it decouples its operation from the infected host's IP address, the traffic that is carried does not leak information about the source, destination and message. In addition, no bot knows the IP of any other bot. To communicate, it only knows a few temporary onion addresses of the bots that it can send a message to next. Therefore, tracking the bot chain is nearly impossible. Through Tor relays, the master can control any bot at any time without revealing its identity.

OnionBot operates in 4 stages. The first, infection, is the phase where unknowing users are infected through phishing spam, remote exploiting, drive-by download, zero-day vulnerability, etc. Once a computer is infected, it enters the rally stage, where it will look for other bots in the Tor network. To do this. It bootstraps into the network through a hardcoded peer list of onion addresses, which is periodically updated. After connecting to the OnionBot network, the infected computer enters the waiting stage, where it is ready to receive commands from the botmaster through push-based transmissions. Finally, after receiving a command and a target, it enters the execution phase, where it sends out spam or makes requests repeatedly, such as a Distributed Denial of Service (DDoS) attack.~\cite{marchetti2012financial}

OnionBot may seem impossible to take down due to its leveraging of Tor; however, we can cleverly mitigate it using a process called SOAP~\cite{sanatinia2015onionbots}. In SOAP, we first find a bot's onion address by detecting it through reverse engineering an infected host, or using honeypots. Once the onion address is found, we set up clone bots that peer with this host. Eventually, with enough sybils, we can dominate the neighbors of the bot and partition it out of the botnet. Doing this repeatedly for the OnionBot will eventually neutralize it.

There is no threat of OnionBot at the moment, as it is just a conceptual design implemented in paper of Sanatinia et al. However, there is a very real chance that Tor-based botnets will look sometime like the OnionBot in the future, which would be very tough to mitigate in a timely manner. As more botnet implementers are turning towards monetizing their systems, e.g. botnet-as-a-service, the motivation for creating an effective botnet is higher than ever.

\section{Broken Cryptographic Mechanisms}
Although the increasing use of privacy-enhancing systems by antagonistic software makes the antagonistic programs hard to detect, the primary concern with the ever-increasing sophistication of such malicious systems does not only lie with its potential to distribute malware on end hosts, but also its power to subvert and exploit existing cryptographic mechanisms that many applications utilize each day.

Aside from the typical Internet traffic, it is important to remember that essentially all existing data, such as those needed by anti-virus systems, make use of these cryptographic mechanisms to securely deliver the latest malware signatures in order to operate properly. Without the proper malware signatures, the majority of anti-virus systems would not be able to detect subsequent intrusions by the botnets. In such a serious case, even firewalls would be rendered useless. Once the next wave of botnets successfully exploit the cryptographic mechanisms, not only would it be a privacy concern, but general network security would also be a devastating problem.

\section{Conclusion}
While cryptographic mechanisms and standardized security protocols have come a long way to detect and defend against eavesdroppers using a variety of encryption techniques and key-exchange methods, and Tor that further enhances privacy, malicious systems that couple botnets with Tor, like OnionBot, can lead to serious security concerns. Especially with users already using Tor to enhance their privacy, OnionBot, which operates within Tor, can have the potential to easily infect the connected Tor users. Users who utilize Tor for private communications could one day have their privacy eroded without noticing. Furthermore, these antagonistic programs, more sophisticated and hard to detect than ever, can launch malware that subvert the privacy of many Internet-connected users before users even have a chance to route messages through secure channels.

As new systems like OnionBot gain popularity among the hacker community, existing privacy-enhancing techniques will become evidently less secure, and new security infrastructure will need to be developed. Additional systems similar to Tor will also need to be developed as Tor is already a high-value target for hackers and intelligence agencies. Right now, only very few complete alternatives to Tor exists.

At present, the technologies in use on modern networks, assuming a strong and proper implementation, are still largely safe. However, the developments of new malicious technologies with the potential to subvert existing network security protocols should not be taken lightly. Threats breaking out on privacy enhancing infrastructures can have a serious effect on society and mitigation techniques need to continue evolving.

\bibliographystyle{plain}
\bibliography{CS4740_ProblemSet5}

\begin{thebibliography}{10}

\bibitem{christin2012silkroad}
Nicolas Christin.
\newblock Traveling the silk road: A measurement analysis of a large anonymous
  online marketplace.
\newblock In {\em Proceedings of the 22Nd International Conference on World
  Wide Web}, 2012.

\bibitem{gu2008botminer}
Guofei Gu, Roberto Perdisci, and Wenke Lee.
\newblock Botminer: Clustering analysis of network traffic for protocol- and
  structure-independent botnet detection.
\newblock In {\em USENIX Security Symposium}, 2008.

\bibitem{keith2013cryptolocker}
Keith Jarvis.
\newblock Cryptolocker ransomware.
\newblock
  http://www.secureworks.com/cyber-threat-intelligence/threats/cryptolocker-ransomware,
  December 2013.
\newblock Accessed: 2015-12-19.

\bibitem{moshchuk2009spammingBotnets}
John~P. John, Alexander Moshchuk, Steven~D. Gribble, and Arvind Krishnamurthy.
\newblock Studying spamming botnets using botlab.
\newblock In {\em Proceedings of the 6th USENIX Symposium on Networked
  Systems}, 2009.

\bibitem{marchetti2012financial}
Mirco Marchetti, Michele Colajanni, Michele Messori, Leonardo Aniello, and Ymir
  Vigfusson.
\newblock Cyber attacks on financial critical infrastructures.
\newblock In {\em Collaborative Financial Infrastructure Protection}, pages
  53--82. Springer Berlin Heidelberg, 2012.

\bibitem{perdisci2015malwareBehavior}
Roberto Perdisci, Wenke Lee, and Nick Feamster.
\newblock Behavioral clustering of http-based malware and signature generation
  using malicious network traces.
\newblock In {\em Proceedings of the 7th USENIX Conference on Networked Systems
  Design and Implementation}, 2010.

\bibitem{gomez1993taxonomyBotnet}
Rafael~A. Rodriguez-Gomez, Gabriel Macia-Fernandez, and Pedro Garcia-Teodoro.
\newblock Survey and taxonomy of botnet research through life-cycle.
\newblock {\em ACM Comput. Surv.}, 2013.

\bibitem{sanatinia2015onionbots}
Amirali Sanatinia and Guevara Noubir.
\newblock Onionbots: Subverting privacy infrastructure for cyber attacks.
\newblock In {\em The Annual IEEE/IFIP International Conference on Dependable
  Systems and Networks (DSN)}, 2015.

\bibitem{tarakanov2013zeus}
Dmitry Tarakanov.
\newblock The inevitable move – 64-bit zeus enhanced with tor.
\newblock
  http://securelist.com/blog/events/58184/the-inevitable-move-64-bit-zeus-enhanced-with-tor,
  December 2013.
\newblock Accessed: 2015-12-19.

\bibitem{2012databreach}
Verizon~RISK Team.
\newblock 2012 data breach investigation report.
\newblock
  http://www.verizonenterprise.com/resources/reports/rp\_data-breach-investigations-report-2012-ebk\_en\_xg.pdf,
  December 2012.
\newblock Accessed: 2015-12-19.

\bibitem{yadav2010algorithmicallyGeneratedDN}
Sandeep Yadav, Ashwath~K.K. Reddy, and A.L.~Narasimha Reddy.
\newblock Detecting algorithmically generated malicious domain names.
\newblock In {\em Proceedings of the 10th ACM SIGCOMM conference on Internet
  measurement}, 2010.

\end{thebibliography}

\end{document}